\documentclass[review,12pt]{elsarticle}

\usepackage{graphicx}

\usepackage{amssymb}

\usepackage{lineno}

\journal{Journal of Alloys and Compounds}

\begin{document}

\begin{frontmatter}



\title{Influence of substitutional disorder on the
electrical transport and the superconducting properties of
Fe$_{1+z}$Te$_{1-x-y}$Se$_{x}$S$_{y}$}


\author[label1]{M. G. Rodr\'{\i}guez \footnote{Scholarship from CONICET of Argentina}}
\author[label2]{G. Polla}
\author[label2]{C. P. Ramos}
\author[label1]{C. Acha \footnote{Corresponding author: acha@df.uba.ar}}

\address[label1]{Laboratorio de Bajas Temperaturas, Departamento de F\'{\i}sica, FCEyN, UBA and IFIBA-CONICET, Buenos Aires,
  Argentina}
\address[label2]{Gerencia de Investigaci\'on y Aplicaciones, CAC-CNEA, San Mart\'{\i}n, Argentina}

\begin{abstract}

We have carried out an investigation of the structural, magnetic,
transport and superconducting properties of
Fe$_{1+z}$Te$_{1-x-y}$Se$_x$S$_y$ ceramic compounds, for $z=0$ and
some specific Se (0$\leq$ x $\leq$ 0.5) and S (0 $\leq$ y
$\leq$0.12) contents. The incorporation of Se and S to the FeTe
structure produces a progressive reduction of the crystallographic
parameters as well as different degrees of structural disorder
associated with the differences of the ionic radius of the
substituting cations. In the present study, we measure transport
properties of this family of compounds and we show the direct
influence of disorder in the normal and superconductor states. We
notice that the structural disorder correlates with a variable range
hopping conducting regime observed at temperatures $T >$ 200 K. At
lower temperatures, all the samples except the one with the highest
degree of disorder show a crossover to a metallic-like regime,
probably related to the transport of resilient-quasi-particles
associated with the proximity of a Fermi liquid state at
temperatures below the superconducting transition. Moreover, the
superconducting properties are depressed only for that particular
sample, in accordance to the condition that superconductivity is
affected by disorder when the electronic localization length $\xi_L$
becomes smaller than the coherence length $\xi_{SC}$.

\end{abstract}


\end{frontmatter}


\section{Introduction}

Since the discovery of superconductivity in the iron pnictide
system~\cite{Kamihara08}, ReFeAsO (Re: rare earth) doped by chemical
substitution on the Re or in the O site~\cite{Ren08,Hunte08}, at
temperatures between 40 K and 55 K and with critical fields of 65 T
at 4 K, a large quantity of studies was devoted to these Fe-based
superconductors. Among them, it was shown~\cite{Hsu08} the existence
of a related family of superconductors, the iron chalcogenides,
corresponding to FeSe and FeTe and their partial substitutions
FeTe$_{1-x}$Se$_x$. This system has the simplest structure among the
Fe-based superconductors because it is formed just with two layers
of Fe and Se/Te, with an anti-PbO structure (space group
P4/nmm)~\cite{Cryst96}.

The FeTe compound exhibits antiferromagnetic (AF) ordering below
$T_N \sim$70 K and no superconducting state~\cite{Fang08} down to 4
K. By partial substitution of Te by Se or by S, the AF order is
suppressed and superconductivity develops~\cite{Yeh08,Mizu09} at
critical temperatures $T_c \sim$ 8 K - 14 K. On the other hand, FeSe
is a superconductor and the partial substitution of Se by S
initially favors superconductivity, while for replacements over a
20\% superconductivity is depressed~\cite{Mizu09b}. Up to now,
studies of the influence of disorder on the properties of iron
chalcogenides were particularly devoted to substitutions into the
Fe-site~\cite{Nabeshima12,Bezusyy14}. Here, in order to gain insight
on the influence of structural parameters on the normal state
properties and on the appearance of the superconducting state on
these Fe-based chalcogenide superconductors, we have synthesized new
ceramic samples, performing partial substitutions of Te in the FeTe
compound by both Se and S, corresponding to
Fe$_{1+z}$Te$_{1-x-y}$Se$_x$S$_y$ (FTSeS). The Se and S
concentrations were chosen in order to maintain nearly constant the
ionic radius while producing a small increase of the structural
disorder in the chalcogenide site.

In the present work, we particularly report on the synthesis and
structural characterization of the FTSeS ceramic samples and on
their magnetic, transport and superconducting properties. We found
that the structural disorder of the chalcogenide site ($\sigma$)
influences the transport properties, determining a 3D-variable range
hopping conduction-regime observed for temperatures $T >$ 200 K,
with an electronic localization length ($\xi_L$) that decreases
concomitantly with increasing $\sigma$. We found a marked reduction
of $T_c$ for the sample where $\xi_L$ becomes lower that the
coherence length ($\xi_{SC}$), indicating the sensibility of the
superconducting state to nonmagnetic disorder.

\section{Experimental Details}

We have synthesized, by solid-state reaction, the family of FTSeS
ceramics compounds, with nominal compositions z=0 and (x, y) = (0,
0.12); (0.10, 0.10); (0.25, 0.06); (0.30, 0.05); (0.40, 0.02);
(0.45, 0.01) and (0.50, 0). The x and y values were chosen in order
to produce a small variation of the average chalcogenide ionic
radius ($<R_{Ch}> = \sum x_iR_i $, where the $x_i$ are the
fractional occupancies of each chalcogenide and $R_i$ their ionic
radius) while increasing the random disorder associated with their
different ionic radius. To vary this disorder, we use different Te,
Se and S combinations in order to modify the standard deviation
$\sigma = \sqrt{(\sum x_iR_i^2-<R_{Ch}>^2)}$ and $<R_{Ch}>$ was
chosen around (2.12 $\pm$ 0.04) \AA. As the starting materials, we
used high-purity elements, Fe, Te, Se and S (Alfa Caesar 99.99 \%).
Considering their increasing Se contents these samples were labeled
S01 to S07, respectively.  The materials were ground, pelletized and
encapsulated in a quartz tube in vacuum. Then, the samples were
heated at 750 $^{\circ}$C for over 24 hours and slowly cooled down
to room temperature. We repeated the heat treatments again and
finally, we resintered the samples in a rectangular shape at 750
$^{\circ}$C but reducing time to 12 hours.

Samples were characterized by x-ray powder diffraction (XRD) using a
CuK$\alpha$ radiation with a PANalytical Empyrean model
diffractometer with PIXcel 3D detector. A complete Rietveld
refinement was done with Full Prof program~\cite{Carv93}.
M\"ossbauer spectra of the S07 sample were obtained at room
temperature (RT) and 20 K with a conventional constant acceleration
spectrometer in transmission geometry with a $^{57}$Co/Rh source.
Measurements were recorded at 4 mm/s and 11 mm/s and then fitted
using the Normos program~\cite{Brand87}. Isomer shift values are
given relative to that of $\alpha$-Fe at RT. A quantitative point
analysis was performed by energy dispersive x-ray spectroscopy (EDX)
coupled with a scanning electron microscopy (SEM). All the samples
were polished before the measurements. We made several measurements
on the whole surface and then we averaged to obtain a more realistic
value of content of each sample. Electrical resistivity measurements
as a function of temperature (4 K $\leq$ T $\leq$ 300K) were
performed by a four-point method. Magnetization as a function of
temperature and hysteresis loops were achieved using a squid
magnetometer.

\section{Results and Discussion}
\begin{figure} [h]
\vspace{3mm}
\centerline{\includegraphics[angle=0,scale=0.35]{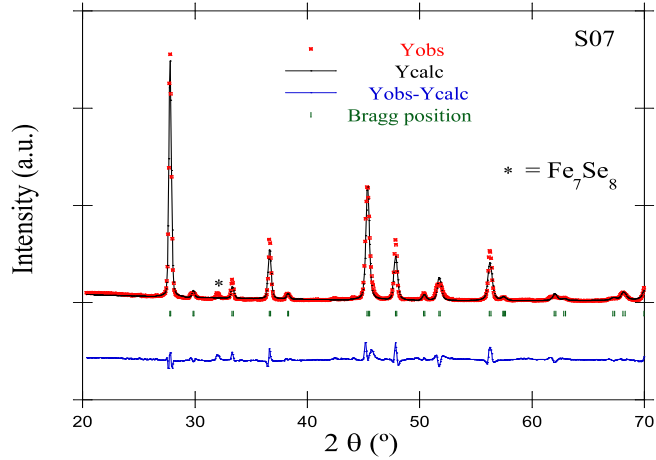}}
\vspace{-5mm}\caption{(Color online) XRD paterns for sample S07 (red
squares) and its Rietveld refinement (black line). Differences and
the Bragg position of the reflexions correspond to blue lines and to
green vertical lines, correspondingly. The presence of the
Fe$_7$Se$_8$ impurity is also indicated.} \vspace{-0mm}
\label{fig:RayosX}
\end{figure}

A typical XRD pattern obtained in this case for sample S07 and its
Rietveld refinement can be observed in Fig.~\ref{fig:RayosX}.
Similar results were obtained for the whole batch of FTSeS samples.
The refinements were performed by considering a tetragonal structure
P4/nmm. The presence of a fraction less than $\sim 8\%$ of
Fe$_7$Se$_8$ as impurity was also refined for all the samples. Peaks
associated with this impurity were marked with an asterisk in
Fig.~\ref{fig:RayosX}.

\begin{figure} [h]
\vspace{3mm}
\centerline{\includegraphics[angle=0,scale=0.5]{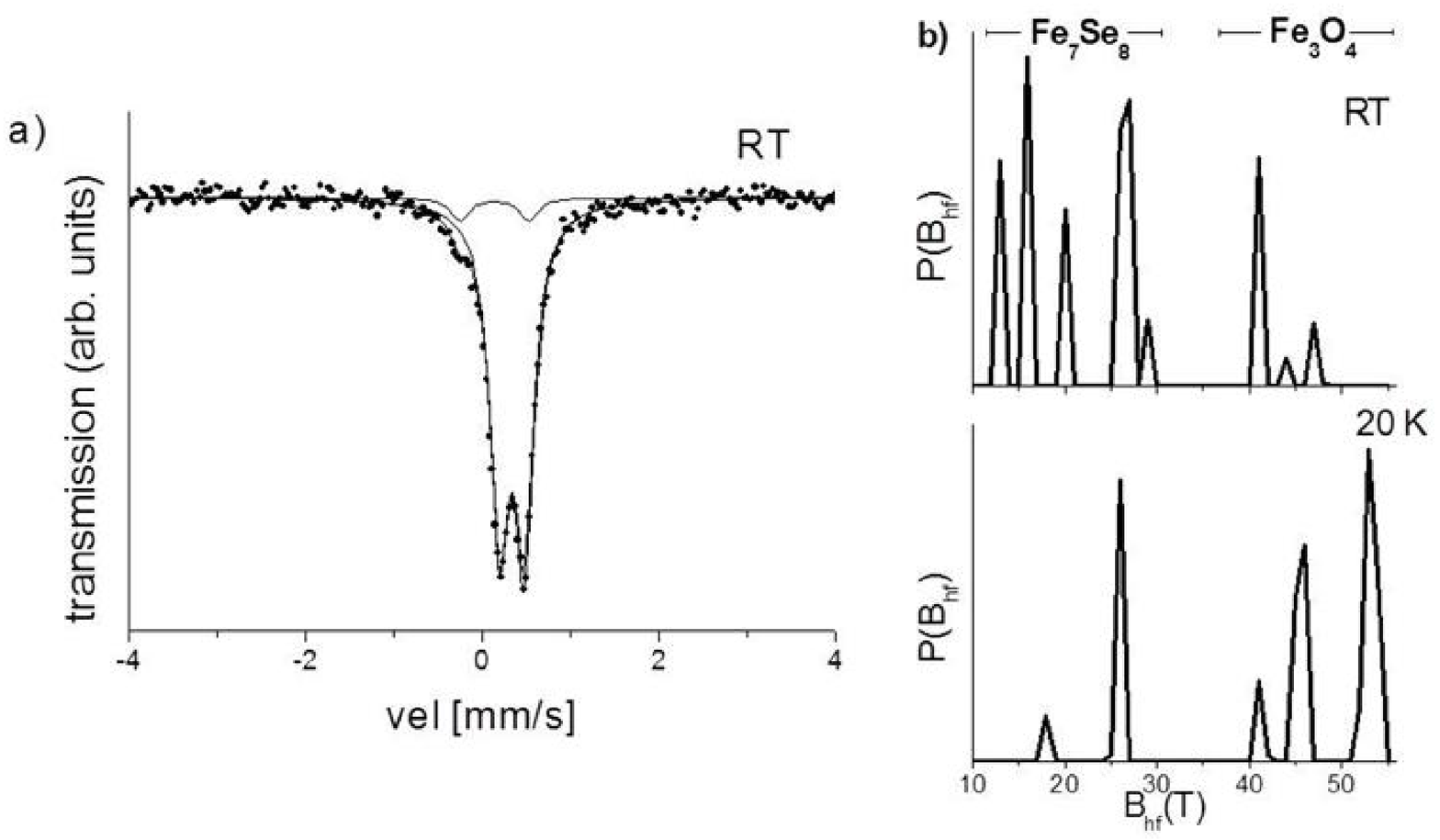}}
\vspace{-5mm}\caption{a) M\"ossbauer spectrum of
FeTe$_{0.5}$Se$_{0.5}$ in a narrow velocity range at RT, including
paramagnetic components fitting, b) Hyperfine field distributions at
RT and 20 K for the magnetically ordered contributions.}
\vspace{-0mm} \label{fig:Moss}
\end{figure}

For a more detailed characterization of sample S07, a M\"ossbauer
spectrum was recorded at room temperature (RT) and 11 mm/s (not
shown here). A dominant broad doublet showed up and a minor content
of impurity magnetic phases seemed to be hidden in the background.
In order to allow a better resolution of the paramagnetic hyperfine
splitting, an additional spectrum was recorded at 4 mm/s, in which
two main paramagnetic components are clearly distinguished
(Fig.~\ref{fig:Moss}a). The major one has an isomer shift of 0.45
mm/s and a quadrupole splitting of 0.28 mm/s, in coincidence with
the hyperfine parameters reported in literature for
FeTe$_{0.5}$Se$_{0.5}$~\cite{Szy11,Gom10}. The other one instead has
an isomer shift of ~ 0.26 mm/s and a quadrupole splitting of ~ 0.78
mm/s. This last subspectrum with Fe$^{3+}$ character could be due to
a superparamagnetic contribution of impurity phases with particle
sizes typically smaller than 10 nm. Those parameters were then fixed
in the spectrum recorded at RT and 11 mm/s, to take into account
only the minor magnetic component. For that purpose a fitting with a
hyperfine field distribution for the higher fields ($>$ 10 T) was
performed (see Fig.~\ref{fig:Moss}b).

\begin{figure} [h]
\vspace{3mm}
\centerline{\includegraphics[angle=0,scale=0.35]{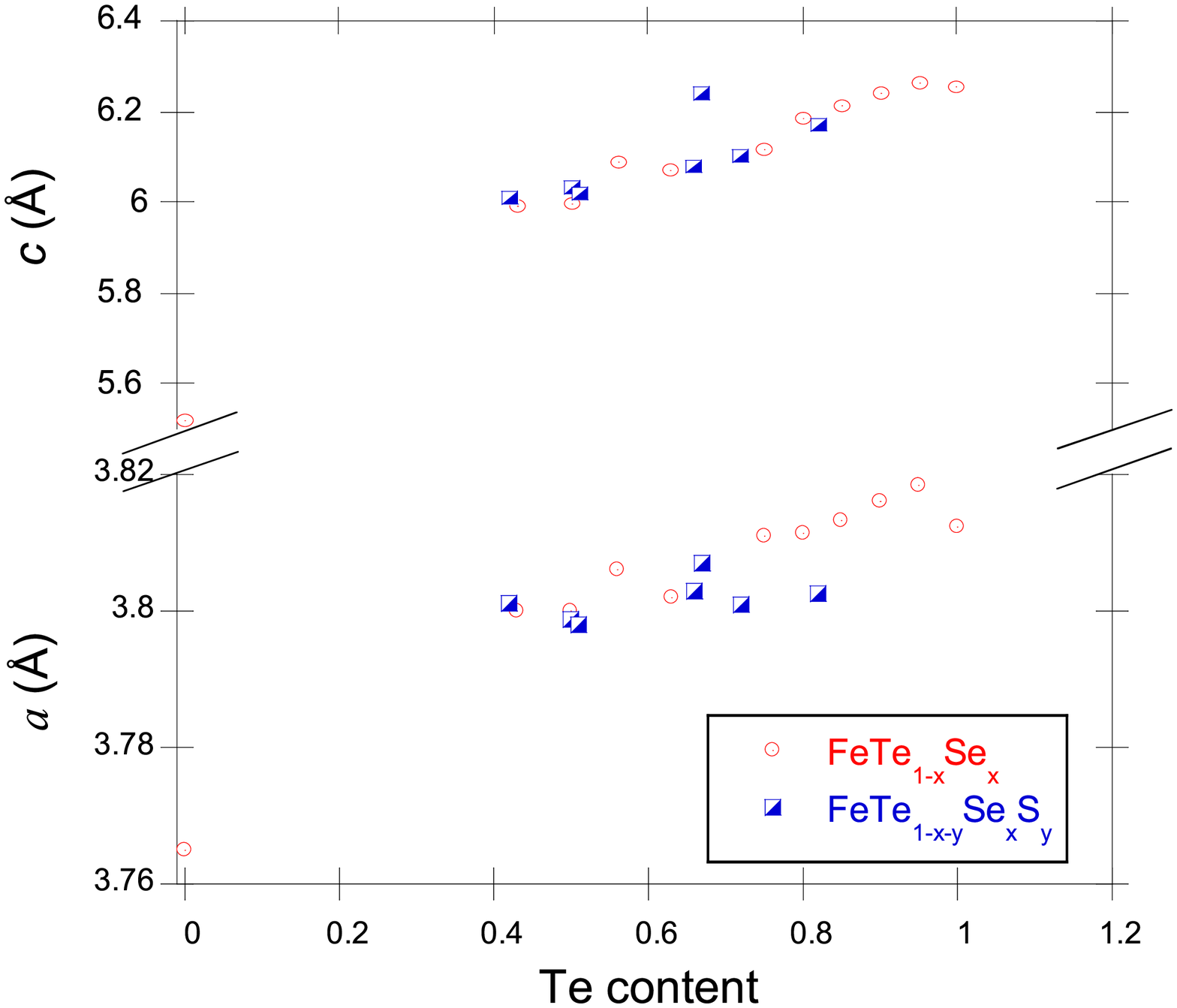}}
\vspace{-5mm}\caption{(Color online) Crystallographic parameters $a$
and $c$ as a function of the real Te content for the FTSeS samples.
Data for the FeTe$_{1-x}$Se$_x$ (FTSe) samples is also included for
comparison (from
ref.~\cite{Marti10,Yeh08,Garba09,Gresty09,Tegel10,Bendele10})}
\vspace{-0mm} \label{fig:aycvsTe}
\end{figure}

This analysis revealed the presence of at least two impurity phases.
An appropriate identification required another spectrum recorded at
low temperature and 11 mm/s. In this case, to ensure a temperature
below any possible transition, 20 K was chosen. The corresponding
hyperfine field distribution is also shown in Fig.~\ref{fig:Moss}b.
The area under the magnetic part of the spectrum slightly increased
when lowering temperature, confirming that as it was supposed above
in the text, the minor doublet contribution at RT arose from
superparamagnetic impurities. Based on the whole analysis, including
RT measurements in wide and narrow velocity ranges and at low
temperature, it can be concluded that one of the impurity phases is
magnetite (Fe$_3$O$_4$); distinguished on cooling due to the Verwey
transition~\cite{Walz02}. Considering the recoilless factor of
Fe$_3$O$_4$ at RT, reported in \cite{Gom10}, the effective fraction
of magnetite in the sample would not exceed 3\%. The other one, a
low field contribution to the hyperfine distribution at RT which
evolves to higher fields as lowering temperature, would agree with
the spin reorientation corresponding to Fe$_7$Se$_8$.~\cite{Ok73}
This impurity fraction represents about 10\% of the iron bearing
compounds in the sample. These results do not contradict the XRD
measurements in which Fe$_7$Se$_8$ with a similar fraction was
detected. Fe$_3$O$_4$ was not distinguished due to the experimental
resolution of the XRD technique (a phase fraction lower than 5\% is
difficult to detect).

\begin{table}[h]
\begin{center}
\caption{Composition determined by EDX ($\pm 0.02$ for Te, Se and S
and $\pm 0.04$ for Fe), standard deviation of the radius of the
chalcogenide site $\sigma$ ($\pm 5\%$) , crystallographic $a$ and
$c$ parameters, superconducting transition temperature defined by
magnetization and transport measurements ($\pm 3\%$), coefficient
T$_0$ (see Eq. 2, $\pm 8\%$) and localization length $\xi_L$ ($\pm
3\%$).}

\begin{small}
\vspace{3mm}
\begin{tabular}{|c|c|c|c|c|c|c|c|c|}
\hline \multicolumn{1}{|l|}{\textbf{Sample}} & \textbf{EDX
composition
 } &  \textbf{$\sigma$} & \textbf{a}  &
\textbf{c} & \textbf{T$_c^{mag}$}  & \textbf{T$_c^{res}$ } &
\textbf{T$_0$ }  & \textbf{$\xi_L$}  \\  & & ({\AA}) &
({\AA}) & ({\AA}) & (K) & (K) & (K) & ({\AA})  \\
\hline \hline
S01 & Fe$_{1.15}$Te$_{0.72}$S$_{0.28}$ & 0.166 &
3.800(9) & 6.10(4) & - & 6.8  & 2294 & 1.8
\\ \hline S02  &
Fe$_{0.88}$Te$_{0.82}$Se$_{0.09}$S$_{0.09}$    & 0.119 & 3.802(5) &
6.16(9) & 6.6 & 9.9 & 403 & 3.3
\\ \hline S03                            &
Fe$_{1.25}$Te$_{0.67}$Se$_{0.27}$S$_{0.06}$     & 0.125 & 3.807(0) &
6.23(8) & 11.3 & 12.3
& 853 & 2.5                                \\
\hline S04                             &
Fe$_{1.18}$Te$_{0.66}$Se$_{0.28}$S$_{0.06}$
  & 0.125 & 3.802(9)
& 6.08(1) & 12.2 & 13.0  & 435 & 3.2                               \\
\hline S05 & Fe$_{0.92}$Te$_{0.50}$Se$_{0.46}$S$_{0.04}$
 & 0.123 & 3.798(7) &
6.03(4) & 13.1 & 13.5 & 337 & 3.5
\\ \hline S06                               &
Fe$_{1.13}$Te$_{0.51}$Se$_{0.47}$S$_{0.02}$
      & 0.120 & 3.798(0)
& 6.01(7) & 13.3 & 13.5     & 297 & 3.6                                \\
\hline S07                               &
Fe$_{0.99}$Te$_{0.42}$Se$_{0.58}$ & 0.114 & 3.801(1) & 6.00(8) &
14.3 & 14.0
& 524 & 3.0                             \\
\hline
\end{tabular}
\end{small}
\end{center}
\end{table}

The evolution of both $a$ and $c$ crystallographic parameters by
increasing the substitution of Te by Se or S can be observed in
Fig.~\ref{fig:aycvsTe}. Our results are compared with those
corresponding to the Te substitution by Se for the
FeTe$_{1-x}$Se$_x$ (FTSe) compound. It can be observed that mainly
the $a$ and $c$ parameters are reduced when decreasing the Te
content of the sample, similarly to the results observed for the
FTSe samples. Although some exceptions can be noted (like an
anomalous high value of the $c$ parameter for sample S03) that can
be associated with an excess of Fe, probably occupying interstitial
sites.~\cite{Pau10}

The real composition of the samples was estimated by the
semi-quantitative determination of an EDX analysis. Five to seven
points were chosen on the surface of the sample in order to get an
average composition. A typical dispersion of 0.02 was obtained for
Te, Se and S, and was slightly higher for Fe (0.04). This higher
value can be related to the presence and distribution of Fe-based
impurities, which may also produce an overestimation of the Fe
content. Results obtained by Rietveld refinements and by EDX
analysis are summarized in Table 1.

\begin{figure} [h]
\vspace{3mm}
\centerline{\includegraphics[angle=0,scale=0.35]{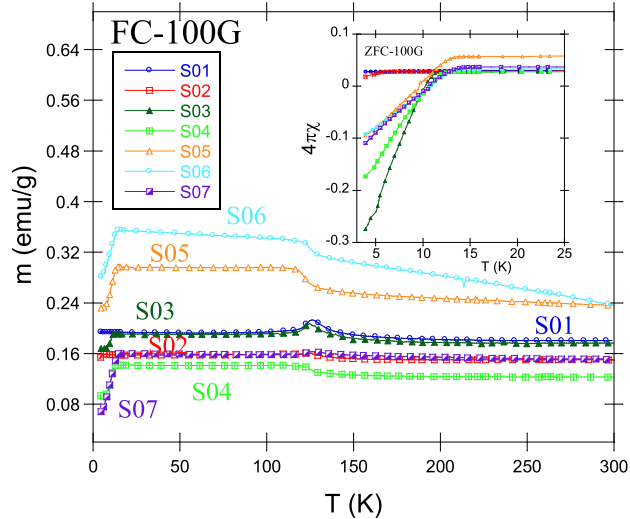}}
\vspace{-5mm}\caption{(Color online) FC magnetization as a function
of temperature for samples S01 to S07 at 100 G. The inset shows the
ZFC DC susceptibility (4$\pi\chi$) as a function of temperature.
T$_c$ can be defined by the position of the onset of the diamagnetic
signal.} \vspace{-0mm} \label{fig:MyMdeTinset}
\end{figure}

Field-cooled (FC) magnetization measurements as a function of
temperature at 100 G for samples S01 to S07 are shown in
Fig.~\ref{fig:MyMdeTinset}. The superconducting transitions for
samples S02 to S07 can be observed at low temperatures while a
ferromagnetic background is also present for all the samples with an
anomalous cusp around 125 K. The inset of Fig.~\ref{fig:MyMdeTinset}
shows the 100 G zero-field-cooled (ZFC) DC susceptibility
measurements (4$\pi\chi$) at low temperatures, which gives a rough
idea of the superconducting volume fraction. The onset and the width
of the superconducting transitions for samples S02 to S07 can be
observed, where the absence of a magnetic shielding for sample S01
can be confirmed. Sample S02 also shows a very small diamagnetic
signal, indicating that superconductivity develops only in a
minority phase. The other samples show a higher superconducting
response, although complete bulk superconductivity is not achieved.
T$_c$ was defined as the onset of this signal and reported in Table
1 as T$_c^{mag}$.

\begin{figure} [h]
\vspace{3mm}
\centerline{\includegraphics[angle=0,scale=0.35]{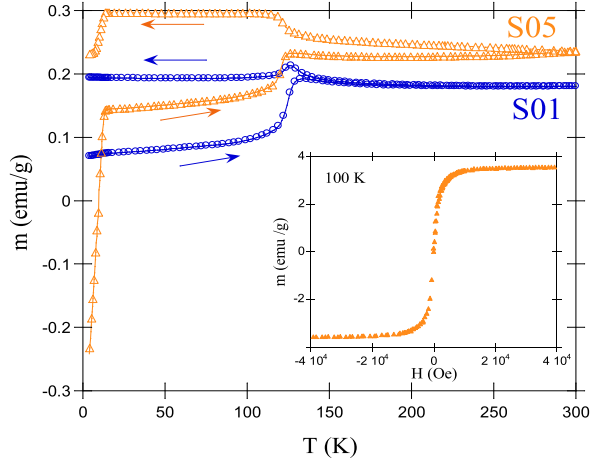}}
\vspace{-5mm}\caption{(Color online) FC and ZFC magnetization (100
G) as a function of temperature for samples S01 and S05. The
observed cusp at $T\simeq 125$ K (Verwey transition) and the low
temperature irreversibilities are characteristic of the presence of
Fe$_3$O$_4$. The inset shows the isothermal magnetization M(H) of
sample S05 at 100 K.} \vspace{-0mm} \label{fig:MdeTyMdeH}
\end{figure}

Fig.~\ref{fig:MdeTyMdeH} shows the FC and the ZFC magnetization
curves for samples S01 and S05. Similar results were obtained for
the other FTSeS samples. The cusp observed around 125 K is the clear
signature of the Verwey transition~\cite{Walz02} associated with the
presence of minor phase Fe$_3$O$_4$. In fact, most of the
magnetization of the normal state can be related to the Fe$_7$Se$_8$
and Fe$_3$O$_4$ impurities~\cite{Bendele10,Kami77}, as the
ferromagnetic background and the hysteretic behavior depicted for
temperatures $T>130 $K is also typical of Fe$_7$Se$_8$, or of
Fe$_3$O$_4$ grains with nanometric
dimensions~\cite{Millo01,Wittlin12}. The magnetization curve M vs H
at T=100 K showed in the inset of Fig.~\ref{fig:MdeTyMdeH}
corresponding to sample S05 (although this curve was practically
sample and temperature independent in the 50 K to 300 K range) is
characteristic of Fe$_3$O$_4$. By considering its saturation value
(3.4 emu/g) we estimate the presence of $\sim$ 4\% in mass of this
impurity~\cite{Goya03}, in accordance with the M\"ossbauer spectrum.

\begin{figure} [h]
\vspace{3mm}
\centerline{\includegraphics[angle=0,scale=0.35]{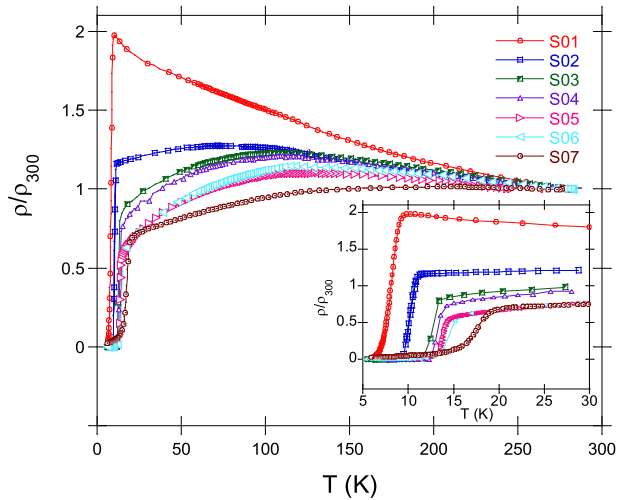}}
\vspace{-5mm}\caption{(Color online) Temperature dependence of the
normalized resistivity ($\rho/\rho_{(300 K)}$) for samples S01 to
S07. The inset shows the detail of the resistive superconducting
transition for samples S01 to S07.} \vspace{-0mm} \label{fig:RdeT2}
\end{figure}

The normalized resistivity ($\rho/\rho_{(300 K)}$) as a function of
temperature is shown in Fig.~\ref{fig:RdeT2}. A semiconducting-like
behavior at temperatures $T > 150$ K can be observed for all the
samples while, for samples S02 to S07, a hump and a metallic-like
conduction is obtained with further decreasing temperature. Only
sample S01 remains as an insulator down to its superconducting
transition. This particular metallic-like conduction observed at low
temperatures, with an increasing slope with decreasing temperature,
is characteristic of many Fe-chalcogenide
superconductors.~\cite{Mizu10} This non-Fermi-Liquid (NFL) behavior
was recently associated with a resilient quasi-particle (RQP)
regime~\cite{Deng13}, where the resistivity is still dominated by FL
quasi-particles, in an intermediate temperature range, between a FL
and a bad metal phase.

In the RQP regime, resistivity is not following the $T^2$ dependence
expected for a FL, nor the linear behavior of a bad metal. This
latter regime is attained when the resistivity reaches the Mott
Ioffe Regel limit ($\rho_{MIR}$).~\cite{Hussey04} Until that point,
resistivity evolves in the RQP regime, showing a negative intercept
and a gradual tendency to a linear regime as the temperature is
increased well above the FL transition temperature.

To check if the observed anomalous metallic-like behavior at low
temperatures (Fig.~\ref{fig:RdeT2}) is developed between these
limits, a rough estimation of $\rho_{MIR}$ can be obtained, assuming
a spherical Fermi surface and a carrier density of
$10^{20}-6~\times10^{20}$ $cm^{-3}$.~\cite{Su13} This gives
$\rho_{MIR} \simeq$ 2-6 m$\Omega$~cm. As the resistivities of our
FTSeS samples (0.2 to 0.4 m$\Omega$~cm) are well below $\rho_{MIR}$,
it is consistent to consider that the observed NFL behavior develops
in a RQP regime.

Unlike what we observed in magnetic measurements, impurities have a
negligible contribution to resistivity. This fact is indeed clear as
at the Verwey transition ($\simeq 125$ K) the resistivity of
Fe$_3$O$_4$ diverges, but no appreciable changes are observed in our
resistivity curves (see Fig.~\ref{fig:RdeT2}).

The inset of Fig.~\ref{fig:RdeT2} shows the detail of the
superconducting transitions, present for all the samples, even for
sample S01, indicating in this case its filamentary nature, as no
superconducting shielding was noticed in the magnetization
measurements down to 4 K (see Fig.~\ref{fig:MyMdeTinset}). We would
like to note here that S01 and S02 samples are those that show a
highly depressed superconducting state (T$_c <$ 10 K, with a
superconducting volume $<$ 1\%), while samples S03 to S07 present a
more robust superconducting phase. We define the superconducting
transition temperature (T$_c$) as the temperature where the
superconducting state percolates ($\rho \rightarrow 0$) and we
reported this value in Table 1 as T$_c^{res}$.

\begin{figure} [h]
\vspace{3mm}
\centerline{\includegraphics[angle=0,scale=0.35]{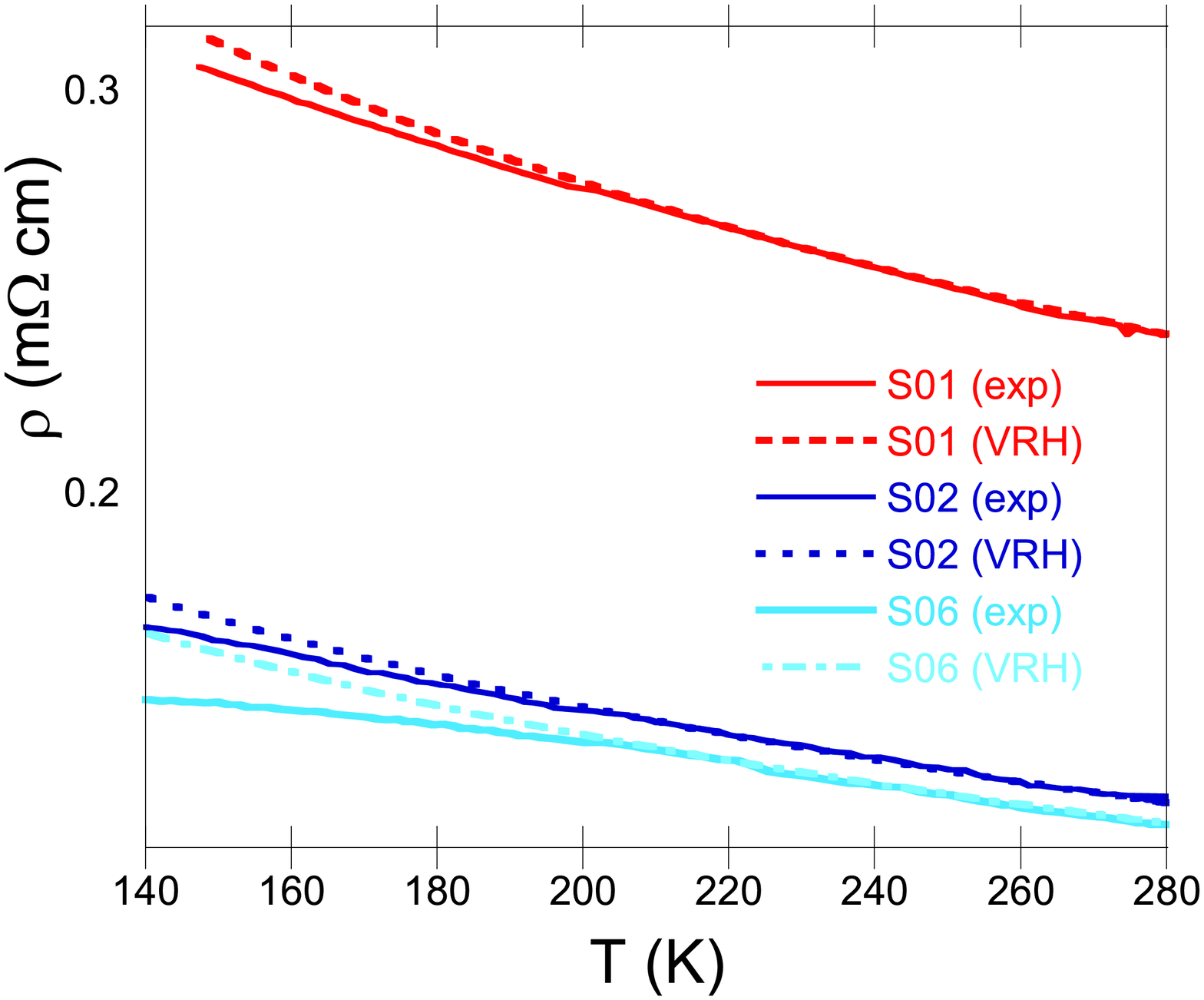}}
\vspace{-5mm}\caption{(Color online) Temperature dependence of
resistivity for samples S01, S02 and S06. A VRH law gives the best
fit of data for temperatures $T>200 K$. Experimental data deviates
from the VRH law for $T \leq$ 200 K.} \vspace{-0mm}
\label{fig:RdeT_VRH}
\end{figure}

We have particularly analyzed the semiconducting-like conduction
regime for the FTSeS samples (150 K $\leq T \leq$ 300 K). To fit the
temperature dependence of the resistivity, we propose a general
expression of the form:

\begin{equation}
\label{eq:rhovsT} \rho = \rho_0~exp[(T_0/T)^n],
\end{equation}

\noindent in order to determine if the conduction regime corresponds
to a semiconductor or to a disordered metal. In the first case,
$T_0$ is associated with the band gap, while in the second case, it
is related to the electronic localization length $\xi_L$, with $T_0$
increasing with increasing disorder~\cite{Efros84}, as expressed in
Eq. 2.

\begin{equation}
\label{eq:localiz} T_0 \simeq \{ \frac{21.2}{k_B N(E_f) \xi_L^3} \},
\end{equation}

\noindent where $k_B$ is the Boltzman constant and $N(E_f)$ the
density of states at the Fermi energy.

\begin{figure} [h]
\vspace{3mm}
\centerline{\includegraphics[angle=0,scale=0.35]{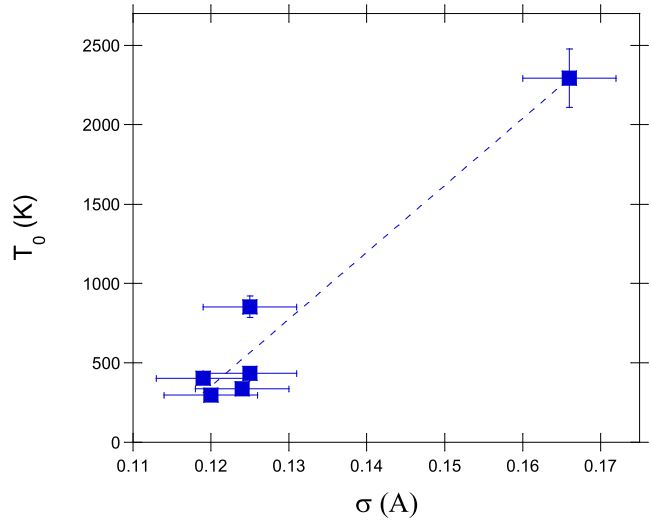}}
\vspace{-5mm}\caption{(Color online) $T_0$ parameter (see Eq.~2) as
a function of the structural degree of disorder $\sigma$.}
\vspace{-0mm} \label{fig:T0vsSigma}
\end{figure}

We found that the best fit is obtained for n=0.25 $\pm$ 0.02, which
agrees with a variable range hopping (VRH) law, indicating that
disorder is determining the electrical conduction of the FTSeS
samples for $T >$ 200 K. Fig.~\ref{fig:RdeT_VRH} shows the results
of these fits for samples S01, S02 and S06. Similar results were
obtained for the other samples. The experimental data is compared
with the VRH law, showing that our samples present an excess
conductivity to the one expected for the VRH law for temperatures $T
<$ 200 K. These deviations from the high temperature VRH conduction
may be associated to a crossover to another conducting regime,
particularly to the RQP mentioned previously. The RQP regime would
dominate the electrical transport at low temperatures, with an
scattering probably associated with incoherent charge dynamics for
samples S02 to S07. The low temperature resistivity of sample S01
could not be fitted by a VRH or a semiconductor temperature
dependence. It is clear that the RQP regime is not developed as for
samples S02 to S07, but it may produce an additional conduction that
impedes to describe the temperature dependence of its resistivity
with a simple expression.

\begin{figure} [h]
\vspace{3mm}
\centerline{\includegraphics[angle=0,scale=0.44]{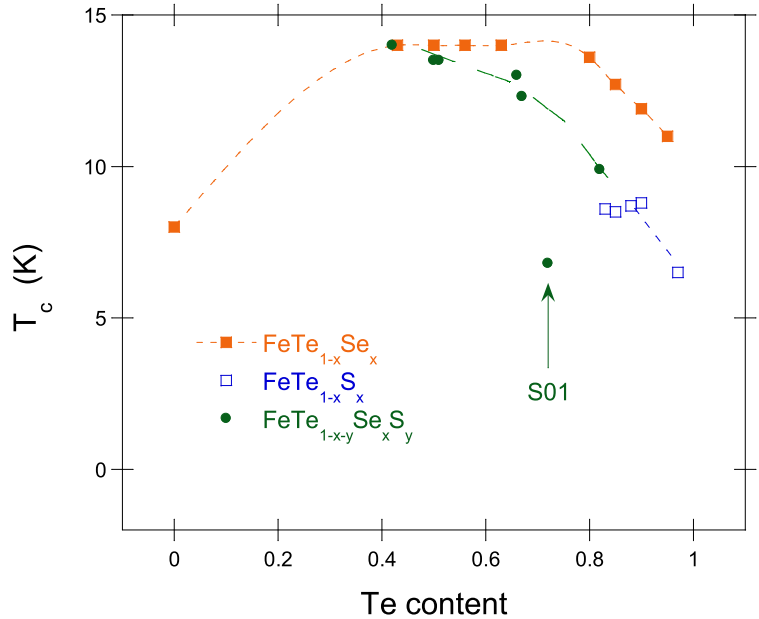}}
\vspace{-5mm}\caption{(Color online) $T_c$ as a function of the Te
content for the family FeTe$_{1-x}$Se$_x$ (FTSe)~\cite{Marti10,
Teg10, Tsu11,Garba09}, FeTe$_{1-x}$S$_x$(FTS)~\cite{Hu09} and FTSeS
(our work). The data point of the S01 sample is indicated with an
arrow. Lines are a guides to the eye.} \vspace{-0mm}
\label{fig:TcvsTe}
\end{figure}

A similar low temperature localization-like regime, with a
filamentary superconducting state was also observed for samples with
high Te content and also extended for lower Te contents by
increasing the interstitial Fe in excess.~\cite{Liu10} Their
electrical transport source of scattering was associated with
dynamic magnetic correlations with in plane magnetic wave vector
($\pi$,0), considered antagonistic to superconductivity.

In Fig.~\ref{fig:T0vsSigma} we have plotted the fitted $T_0$
parameter from the VRH conduction model (see Eq.~1) as a function of
the standard deviation of the chalcogenide radius $\sigma$. Although
the distribution of $\sigma$ values is not uniform in the explored
range, due to the fact that the measured composition differs from
the programmed nominally, the tendency that $T_0$ increases with
increasing $\sigma$ can be conjectured, suggesting that the
structural disorder introduced by substitutions in the chalcogenide
site may be the main source of disorder that controls the electrical
conduction of the normal state.

In order to gain insight in which is the influence of the structural
disorder on the superconducting state, we show in
Fig.~\ref{fig:TcvsTe} the Te content dependence of T$_c^{res}$ for
the families FeTe$_{1-x}$Se$_x$ (FTSe), FeTe$_{1-x}$S$_x$ (FTS) and
FTSeS (our work). It can be observed that the FTSeS samples have an
intermediate $T_c$ between the FTSe and FTS families. Surprisingly,
the $T_c$ of sample S01 is highly depressed in comparison to the
others, clearly out of the general behavior, although its Te
content is not the highest. Its Fe content is also high but it is
even lower than the one of sample S03 which has a higher $T_c$. As
this is the sample with the highest $\sigma$, we may consider if its
weaker and filamentary superconducting state could be a consequence
of its strongly electronic localization. In a simple approach, the
superconducting T$_c$ would be depressed when the localization
length ($\xi_L$) is lower than its superconducting coherence length
($\xi_{SC}$).~\cite{Kotliar86} To verify this possibility, a value
of $\xi_{SC} \simeq$ 2.5 nm was estimated from upper critical field
($H_{c2}$) measurements~\cite{Tarantini11} by assuming a
3D-Ginzburg-Landau relation. The $\xi_L$ of samples FTSeS (see Table
1) can be obtained from Eq.~2 by considering a cell volume $V_c
\simeq $85 \AA, $N(E_f) \simeq 2 States$/eV/$V_c$ ~\cite{Kumar12}
and the temperatures $T_0$ obtained by fitting the VRH conduction
for T$\geq$ 200 K. For sample S01, $\xi_L$ (1.8 nm) is lower than
$\xi_{SC}$, while it is not the case for all the other samples (S02
to S07), in accordance to the assumption than disorder is affecting
its superconducting state.

\section{Conclusions}

To summarize, we have synthesized a new family of superconducting
samples with different degrees of disorder on the chalcogenide site
by partially replacing Te by both Se and S in the FeTe compound. We
determined that magnetic properties in the normal state are
dominated by the presence of small amounts ($\simeq$ 3-4\%) of
Fe$_3$O$_4$ and ($\simeq$ 8-9\%) Fe$_7$Se$_8$, while their influence
on electrical transport properties is negligible. These usual
impurities obtained in the synthesis of the FTSe
compound\cite{Szy11,Wittlin12} produce a ferromagnetic background,
an anomaly near 125 K, corresponding to the Verwey transition of
magnetite, and an hysteretic behavior in the 150 K to 300 K range
probably associated with the ferrimagnetic nature of Fe$_7$Se$_8$.
The AF order, present for the FeTe compound, was not detected for
the range of the substitutions explored. We have shown that disorder
on the chalcogenide site (estimated by the parameter $\sigma$)
influences the electrical resistivity for temperatures higher than
200 K, determining a VRH conduction regime for all the samples. A
metallic-like conduction, with characteristics of a RQP regime,
develops at lower temperatures for all the samples except for the
one with the highest degree of disorder (S01). Particularly for this
sample, superconductivity is highly depressed as expected when
disorder reduces the electronic localization length below the
superconducting coherence length.

\section{Acknowledgments} We would like to acknowledge financial support by CONICET Grant PIP
112-200801-00930 and UBACyT 20020100100679 (2011-2014). We also
acknowledge V. Bekeris for a critical reading, and D. Gim\'enez, E.
P\'erez Wodtke and D. Rodr\'{\i}guez Melgarejo for their technical
assistance.

\section{References}

\bibliographystyle{elsarticle-num}

\begin{thebibliography}{10}
\expandafter\ifx\csname url\endcsname\relax
  \def\url#1{\texttt{#1}}\fi
\expandafter\ifx\csname urlprefix\endcsname\relax\def\urlprefix{URL
}\fi \expandafter\ifx\csname href\endcsname\relax
  \def\href#1#2{#2} \def\path#1{#1}\fi

\bibitem{Kamihara08}
Y.~Kamihara, T.~Watanabe, M.~Hirano, H.~Hosono, J. Am. Chem. Soc.
130 (2008)
  3296.

\bibitem{Ren08}
Z.~Ren, G.~Che, X.~Dong, J.~Yang, W.~Lu, W.~Yi, X.~Shen, Z.~Li,
L.~Sun,
  F.~Zhou, Z.~Zhao, EPL 83 (2008) 17002.

\bibitem{Hunte08}
F.~Hunte, J.~Jaroszynski, A.~Gurevich, D.~C. Larbalestier, R.~Jin,
A.~S. Sefat,
  M.~A. McGuire, B.~C. Sales, D.~K. Christen, D.~Mandrus, Nature 453 (2008)
  903.

\bibitem{Hsu08}
F.~Hsu, J.~Luo, K.~Yeh, T.~Chen, T.~Huang, P.~Wu, Y.~Lee, Y.~Huang,
Y.~Chu,
  D.~Yan, M.~Wu, Proc.Natl. Acad. Sci. U.S.A. 105 (2008) 14262.

\bibitem{Cryst96}
International Tables for Crystallography, Kluwer Acad. Publ. -
Edited by T.
  Hahn, 1996.

\bibitem{Fang08}
M.~H. Fang, H.~M. Pham, B.~Qian, T.~J. Liu, E.~K. Vehstedt, Y.~Liu,
L.~Spinu,
  Z.~Q. Mao, Phys. Rev. B 78 (2008) 224503.

\bibitem{Yeh08}
K.-W. Yeh, T.-W. Huang, Y.~lin Huang, T.-K. Chen, F.-C. Hsu, P.~M.
Wu, Y.-C.
  Lee, Y.-Y. Chu, C.-L. Chen, J.-Y. Luo, D.-C. Yan, M.-K. Wu, EPL (Europhysics
  Letters) 84 (2008) 37002.

\bibitem{Mizu09}
Y.~Mizuguchi, F.~Tomioka, S.~Tsuda, App. Phys. Lett. 94 (2009)
012503.

\bibitem{Mizu09b}
Y.~Mizuguchi, F.~Tomioka, S.~Tsuda, J. Phys. Soc. Jpn. 78 (2009)
074712.

\bibitem{Nabeshima12}
F.~Nabeshima, Y.~Kobayashi, Y.~Imai, I.~Tsukada, A.~Maeda, Effect of
co
  impurities on superconductivity of fese 0.4 te 0.6 single crystals, Japanese
  Journal of Applied Physics 51 (2012) 010102.

\bibitem{Bezusyy14}
V.~Bezusyy, D.~Gawryluk, A.~Malinowski, M.~Berkowski, M.~Cieplak,
Acta Physica
  Polonica 126 (2014) A76.

\bibitem{Carv93}
J.~Rodriguez-Carvajal, Physica B 192 (1993) 55.

\bibitem{Brand87}
R.~A. Brand, Internat. Rep. Angewandte Physik, University of
Duisburg, 1987.

\bibitem{Szy11}
K.~Szyma\'nski, W.~Olszewski, L.~Dobrzy\'nski, D.~Sat\'ua, D.~J.
Gawryluk,
  M.~Berkowski, R.~Pu\'zniak, A.~Wi\'sniewski, Supercond. Sci. Technol. 24
  (2011) 105010.

\bibitem{Gom10}
R.~W. G\'omez, V.~Marquina, J.~L. P\'erez-Mazariego, R.~Escamilla,
R.~Escudero,
  M.~Quintana, J.~J. Hern\'andez-G\'omez, R.~Ridaura, M.~L. Marquina, J.
  Supercond. Nov. Magn. 23 (2010) 551.

\bibitem{Marti10}
A.~Martinelli, A.~Palenzona, M.~Tropeano, C.~Ferdeghini, M.~Putti,
M.~R.
  Cimberle, T.~D. Nguyen, M.~Affronte, C.~Ritter, Phys. Rev. B 81 (2010)
  094115.

\bibitem{Garba09}
G.~Garbarino, A.~Sow, P.~Lejay, A.~Sulpice, P.~Toulemonde,
M.~Mezouar,
  M.~N\'u\~nez Regueiro, EPL 86 (2009) 27001.

\bibitem{Gresty09}
N.~C. Gresty, Y.~Takabayashi, A.~Y. Ganin, M.~T. McDonald, J.~B.
Claridge,
  D.~Giap, Y.~Mizuguchi, Y.~Takano, T.~Kagayama, Y.~Ohishi, M.~Takata, M.~J.
  Rosseinsky, S.~Margadonna, K.~Prassides, Journal of the American Chemical
  Society 131 (2009) 16944--16952.

\bibitem{Tegel10}
M.~Tegel, C.~L$\ddot{o}$hnert, D.~Johrendt, Solid State
Communications 150
  (2010) 383 -- 385.

\bibitem{Bendele10}
M.~Bendele, P.~Babkevich, S.~Katrych, S.~N. Gvasaliya,
E.~Pomjakushina,
  K.~Conder, B.~Roessli, A.~T. Boothroyd, R.~Khasanov, H.~Keller, Phys. Rev. B
  82 (2010) 212504.

\bibitem{Walz02}
F.~Walz, J. Phys. Condens. Matter. 14 (2002) R285.

\bibitem{Ok73}
H.~N. Ok, S.~W. Lee, Phys. Rev. B 8 (1973) 4267--4269.

\bibitem{Pau10}
C.~S. Yadav, P.~L. Paulose, J. Appl. Phys. 107 (2010) 083908.

\bibitem{Kami77}
T.~Kamimura, J. Phys. Soc. Jpn. 43 (1977) 1594.

\bibitem{Millo01}
N.~Guigue-Millot, N.~Keller, P.~Perriat, Phys. Rev. B 64 (2001)
012402.

\bibitem{Wittlin12}
A.~Wittlin, P.~Aleshkevych, H.~Przybylińska, D.~J. Gawryluk,
P.~Dłużewski,
  M.~Berkowski, R.~Puźniak, M.~U. Gutowska, A.~Wiśniewski, Superconductor
  Science and Technology 25 (2012) 065019.

\bibitem{Goya03}
G.~F. Goya, T.~S. Berqu\'o, F.~C. Fonseca, M.~P. Morales, J. Appl.
Phys. 94
  (2003) 3520.

\bibitem{Mizu10}
Y.~Mizuguchi, Y.~Takano, Journal of the Physical Society of Japan 79
(2010)
  102001.

\bibitem{Deng13}
X.~Deng, J.~Mravlje, R.~\ifmmode~\check{Z}\else \v{Z}\fi{}itko,
M.~Ferrero,
  G.~Kotliar, A.~Georges, Phys. Rev. Lett. 110 (2013) 086401.

\bibitem{Hussey04}
N.~E. Hussey, K.~Takenaka, H.~Takagi, Philosophical Magazine 84~(27)
(2004)
  2847.

\bibitem{Su13}
T.~S. Su, Y.~W. Yin, M.~L. Teng, Z.~Z. Gong, M.~J. Zhang, X.~G. Li,
Journal of
  Applied Physics 114 (2013) 183901.

\bibitem{Efros84}
B.~I. Shklovskii, A.~L. Efros, Electronic properties of doped
semiconductors,
  Springer-Verlag, 1984.

\bibitem{Teg10}
M.~Tegel, C.~L\"ohnert, D.~Johrendt, Solid State Comm. 150 (2010)
383.

\bibitem{Tsu11}
V.~Tsurkan, J.~Deisenhofer, A.~G$\ddot{u}$nther, C.~Kant, M.~Klemm,
H.~A.
  Krug~von Nidda, F.~Schrettle, A.~Loidl, Eur. Phys. J. B. 79 (2010) 289.

\bibitem{Hu09}
R.~Hu, E.~S. Bozin, J.~B. Warren, C.~Petrovic, Phys. Rev. B 80
(2009) 214514.

\bibitem{Liu10}
T.~J.~L. et~al., Nature Materials 9 (2010) 718.

\bibitem{Kotliar86}
G.~Kotliar, A.~Kapitulnik, Phys. Rev. B 33 (1986) 3146.

\bibitem{Tarantini11}
C.~Tarantini, A.~Gurevich, J.~Jaroszynski, F.~Balakirev,
E.~Bellingeri,
  I.~Pallecchi, C.~Ferdeghini, B.~Shen, H.~H. Wen, D.~C. Larbalestier, Phys.
  Rev. B 84 (2011) 184522.

\bibitem{Kumar12}
J.~Kumar, S.~Auluck, P.~K. Ahluwalia, V.~P.~S. Awana, Supercond.
Sci. Technol.
  25 (2012) 095002.

\end{thebibliography}

\end{document}